\documentstyle[aps,pre,preprint,amsfonts]{revtex}

\begin{document}

\draft
\tightenlines

\title{Simplicity of State and Overlap Structure in Finite-Volume Realistic Spin Glasses}  
\author{C.M. Newman}
\address{Courant Institute of Mathematical Sciences,
New York University, New York, NY 10012}
\author{D.L. Stein}
\address{Departments of Physics and Mathematics, University of Arizona,
Tucson, AZ 85721}

\maketitle

\begin{abstract}
We present a combination of heuristic and rigorous arguments indicating 
that both the pure state structure 
and the overlap structure of realistic
spin glasses should be relatively simple:  in a large finite volume
with coupling-independent boundary conditions, such as periodic,
at most a pair of flip-related (or the appropriate number of 
symmetry-related in the non-Ising
case) states appear, and the Parisi overlap distribution
correspondingly exhibits at most
a pair of $\delta$-functions at $\pm q_{EA}$.  This rules out the
nonstandard SK picture introduced by us earlier, and when combined
with our previous elimination of more standard versions of the mean field
picture, argues against the possibility of even limited versions of mean field
ordering in realistic spin glasses.  If broken spin flip symmetry should
occur, this leaves open two main possibilities for ordering in the
spin glass phase:  the droplet/scaling two-state picture,
and the chaotic pairs many-state picture introduced by us earlier.  We
present scaling arguments which provide a possible physical basis for
the latter picture, and discuss possible reasons behind numerical
observations of more complicated overlap structures in finite volumes.
\end{abstract}



\section{Introduction}
\label{sec:intro}

Prevalent scenarios \cite{MPV,BY} concerning realistic spin glasses require
that the nature of the spin glass order 
parameter (i.e., the Parisi
overlap distribution) and the structure of the thermodynamic states
from which it is obtained be highly complex; 
see, for example, 
Refs.~\cite{Parisi1,Parisi2,Houghton83,MPSTV1,MPSTV2,ultrametricity,MP,BMY1,BMY2,Parisi3,BCPR,FPV,Ritort94,MPR94,LD,MPRR,MPR97}.
This complexity is asserted to be a consequence of the existence of
many competing pure states.  In previous papers 
\cite{NS96b,NS96c,NS97,NSBerlin,Zurich,NSAustralia} we showed
that the standard picture of this complex structure (including
non-self-averaging of the thermodynamic overlap distribution
function, ultrametricity of distances among all pure states
for fixed coupling realization, etc.) cannot hold in any finite dimension.
However, at the same time we presented (as a logical
possibility) a {\it nonstandard\/} mean field 
picture in which some of these features appear in finite-dimensional
spin glasses but in a more limited sense --- i.e., in large finite volumes
with coupling-independent boundary conditions such as periodic.  In this
picture,  only a {\it subset\/} of all the pure states appears 
in each finite-volume mixed
state (which varies with volume); those pure states along with
their weights and overlaps retain some mean field structure.

In this paper, however, we provide both heuristic 
and rigorous arguments that indicate the state and overlap structure in
finite volumes must in fact be relatively simple.  This is so
even if there are many pure states overall.  These 
arguments preclude the possibility
of any type of mean field structure --- even the nonstandard, limited type
--- for the spin glass phase in finite dimensions. 

Although the arguments and conclusions of this paper are applicable to fairly
general examples of disordered systems, 
we will focus on the Edwards-Anderson (EA)
Ising spin glass \cite{EA}. 
When there are many pure (infinite volume) states $\rho^\alpha$,
it has been generally believed \cite{MPV} that the finite volume Gibbs state
$\rho^L_{\cal J}$ (for a coupling configuration 
${\cal J}$ in the cube $\Lambda_L$ of side $L$ 
centered at the origin with, say,
periodic boundary conditions) is (approximately) a mixture of many
pure states:
\begin{equation}
\label{eq:sum}
\rho_{\cal J}^{(L)}\approx\sum_\alpha 
W_{{\cal J},L}^\alpha\rho_{\cal J}^\alpha 
\end{equation}
and the finite volume overlap distribution $P_{\cal J}^L(q)$
is (approximately) the corresponding 
mixture of many $\delta$-functions:
\begin{equation}
\label{eq:PL}
P_{\cal J}^L(q)\approx\sum_{\alpha,\gamma}W_{{\cal J},L}^\alpha 
W_{{\cal J},L}^\gamma
\delta(q-q_{\cal J}^{\alpha\gamma})\quad ,
\end{equation}
where $q_{\cal J}^{\alpha\gamma}$ is the overlap between the pure
states $\alpha$ and $\gamma$:
\begin{equation}
\label{eq:qab}
q_{\cal J}^{\alpha\gamma} = \lim_{L' \to \infty}
|\Lambda_{L'}|^{-1}\sum_{x\in\Lambda_{L'}}\langle\sigma_x\rangle^\alpha
\langle\sigma_x\rangle^\gamma \quad .
\end{equation}

Of course, if there is only a single pair of pure states (related by a global spin
flip) as in the droplet/scaling picture of
Refs.~\cite{FH86,HF87a,HF87b,FH88}
(see also \cite{Mac,BM}), 
then for each $L$, $P_{\cal J}^L(q)$ will simply 
approximate a sum of two  $\delta$-functions
at $\pm q_{EA}$.  We will argue here that {\it the same conclusion is true for the
finite-volume overlap distributions even when there are many pure
states\/}.  This is because $\rho^L_{\cal J}$ will still be approximately 
a mixture of a {\it single\/} pair of pure states, although now the choice 
of the pair will depend upon $L$. This
scenario was previously proposed in Refs.~\cite{NS96c,NS97,NSBerlin} as a logical possibility
that followed from the metastate approach introduced in those papers.
Here we will argue that it is the only {\it reasonable\/}
possibility consistent
with many pure states, and we will also present new scaling arguments that
provide a physical basis for it and at the same time explain its relation
to the earlier and simpler two state droplet/scaling picture.

It is important to note that in computing overlap distributions as
in Eq.~(\ref{eq:PL}), the region in which the computation is done should
be small compared to the overall size of the system --- i.e., the
system boundaries should be far from 
the region of interest.  The reasons for this
were discussed at some length in the Appendix to Ref.~\cite{NS97},
and will be returned to in Section~\ref{sec:pure}.  
This guarantees that one is
focusing on the thermodynamic states of the system \cite{NS97,HF87a}
and avoiding extraneous finite size and boundary effects.

With this understanding, our arguments indicate that the nonstandard
SK picture, introduced by us previously as the only remaining viable
mean-field-like picture, is not valid in any dimension.  The reader
may wish to glance ahead at Section~\ref{sec:invariance} in which
this conclusion, one of the main results of the paper, is presented.

The plan of the paper is as follows.  In
Section~\ref{sec:meta} we review the concept of metastates.  In
Section~\ref{sec:finite} we discuss previously 
proposed scenarios for the spin glass
phase, including the newer chaotic pairs and nonstandard SK pictures.  In
Section~\ref{sec:invariance} we present the first of our
main results, a theorem on the invariance of the metastate
with respect to flip-related boundary conditions, and then
discuss the consequences of the theorem. 
We will see why this result should be incompatible with any but the
simplest spin glass ordering, and in particular how that argues against
the nonstandard SK picture.  In Section~\ref{sec:scaling} we will
provide a scaling basis for the chaotic pairs picture, and present
one possible physical scenario under which it would occur.
In Section~\ref{sec:pure} we discuss, in light of our
results, the question of why some numerical experiments
appear to see a complicated overlap structure.  We further discuss 
appropriate procedures for computing overlap structures in finite volumes
as a means of extracting at least partial information on ordering
in the low temperature phase.  Finally, in Section~\ref{sec:conclusions}
we present our conclusions.

\section{Metastates}
\label{sec:meta}

For specificity, we will focus on the Edwards-Anderson (EA) model \cite{EA}
which, on a cubic lattice
in $d$ dimensions, is described by the Hamiltonian
\begin{equation}
\label{eq:EA}
{\cal H}_{\cal J}(\sigma)= -\sum_{\langle x,y\rangle} J_{xy} \sigma_x \sigma_y\quad ,
\end{equation}
where ${\cal J}$ denotes the set of couplings $J_{xy}$ and where
the brackets indicate that the sum is over nearest-neighbor
pairs only, with the sites $x,y\in Z^d$.  We will take the spins
$\sigma_{x}$ to be Ising, i.e., $\sigma_{x}=\pm 1$;
although this will affect the details of
our discussion, it is unimportant for our main conclusions.
The couplings $J_{xy}$ are quenched, independent, identically
distributed random variables; throughout the paper we will
assume their common distribution to be symmetric about zero
(and usually with the variance fixed to be one).
The most common examples are the
Gaussian and $\pm J$ distributions.
The infinite-ranged version of the EA model was introduced
by Sherrington and Kirkpatrick \cite{SK} and is commonly
referred to as the SK model.

Numerical studies of spin glass overlap structure in the EA model study finite
volume cubes with (usually) periodic 
boundary conditions \cite{MPR97,RBY,Caracciolo}.
A crucial property of disordered systems with many competing states is that,
although particular pure states
may be picked out by a special choice of boundary conditions
depending on the disorder realization, such boundary conditions
are not relevant for comparison to either experiments on physical
spin glasses or to numerical simulations.  In all of these cases
boundary conditions are chosen {\it independently\/} of the coupling
realization.  

In this paper we will therefore focus on either fixed or periodic
boundary conditions (and their flip-related b.c.'s; see
Section~\ref{sec:invariance}) 
chosen independently of the couplings.  From a theoretical point of view, 
observable properties in this situation are amenable to analysis 
by means of the metastate 
approach \cite{NS96c,NS97,NSBerlin,Zurich,NSAustralia}.

Metastates enable us to relate the observed 
behavior of a system in large but finite volumes with
its thermodynamic properties.  This relation is relatively straightforward
for systems with few pure states or for those whose states are
related by well-understood symmetry transformations; but
in the presence of many pure states not related by any
obvious transformations, this relation may be subtle
and complex.  In these cases the metastate approach may be
highly useful.

One reason for this is that, in the presence of many competing
pure states, a sequence as $L\to\infty$ of finite-volume Gibbs measures on
cubes $\Lambda_L$ with coupling-independent b.c.'s will generally {\it not\/}
converge to a single limiting thermodynamic state \cite{NS92}.
We call this phenomenon {\it chaotic size dependence\/} (CSD).
In the metastate approach, we exploit 
the presence of CSD by replacing the study of
single thermodynamic states (as is conventionally
done) with an {\it ensemble\/} of 
(pure or mixed) thermodynamic states.  This approach, 
based on an analogy to chaotic dynamical systems, enables
us to construct a limiting measure, but 
it is a measure on the thermodynamic
states themselves.  

This (infinite-volume) measure contains far more
information than any single thermodynamic state.
It has a particular usefulness in the context
of the study of finite volumes because it carries --- among
other things --- the following information.  Suppose 
that there exist many thermodynamic states in some
(fixed) dimension and at some (fixed) temperature.
Then (for example) the periodic b.c.~metastate (constructed
from an infinite sequence of finite-volume Gibbs measures on cubes
with periodic boundary conditions) tells us the likelihood of
appearance of any specified thermodynamic state, pure or mixed,
in a typical large volume.  More precisely, it provides
a probability measure for all possible $1,2,\ldots,n$-point correlation
functions contained in a box centered at the origin whose sides are
sufficiently far from any of the boundaries so that finite size effects
do not appreciably affect the result.

Details on the construction and properties of the metastate
were given in previous papers \cite{NS96c,NS97,NSBerlin}.  Here we simply recount
some central features.  The histogram, or empirical distribution
approach, is a type of microcanonical ensemble which
considers at fixed ${\cal J}$ a sequence of volumes
with specified b.c., such as periodic.  The resulting
sequence of finite-volume Gibbs states $\rho_{\cal J}^{(L_1)},
\rho_{\cal J}^{(L_2)},\ldots,\rho_{\cal J}^{(L_N)}$ each is given
weight $N^{-1}$.  This ``histogram'' of finite-volume Gibbs
states converges to some $\kappa_{\cal J}$ as $N\to\infty$.
The (periodic b.c., in this example) metastate $\kappa_{\cal J}$ is a probability measure
on thermodynamic states $\Gamma$ at fixed ${\cal J}$, and specifies the
fraction of cube sizes that the system spends in each different 
(possibly mixed) thermodynamic state $\Gamma$ \cite{noteinterface}. 

An alternative (and earlier) construction of the metastate, in which
the randomness of the couplings is used directly
to generate an ensemble of states, was provided
by Aizenman and Wehr \cite{AW}.  In this approach one considers
the limiting joint distribution $\mu({\cal J},\rho_{\cal J}^{(L)})$ as
$L\to\infty$.  Technical details can be found in \cite{NS97,NSBerlin,Zurich,AW}.

It can be proved that there exists at least a ${\cal J}$-independent
subsequence of volumes along which the two approaches (empirical
distribution and Aizenman-Wehr) yield the {\it same\/} limiting metastate
\cite{NS97,NSBerlin,Zurich}.  This will be important in what follows 
\cite{subsequence}.

Occasionally a distinction is drawn between
finite- and infinite-volume states (see, for example, \cite{MPR97}),
where it is argued that the first is more physical and the second
merely mathematical in nature.  While we have shown \cite{NS97} 
that the relation between the two may be
more subtle than previously realized --- at least in the case
where many competing states are present --- we also argue that
the distinction drawn above is misleading.  Indeed, it should
be clear from the discussion above that the 
metastate approach is specifically constructed to consider
both finite and infinite volumes together and to unify
the two cases.  In the next section, guided
by this approach, we review 
various allowable scenarios for the EA spin
glass phase.

\section{The Finite-Dimensional Spin Glass Phase}
\label{sec:finite}

Of the possible scenarios for spin glasses at low temperature,
the simplest is that spin-flip 
symmetry is not broken at positive temperatures
in any dimension.  This would be the case if there were no phase
transition at all and the paramagnetic state persisted to arbitrarily
low temperatures.  It would also be the case if there {\it were\/}
a phase transition but the EA order parameter $q_{EA}$ (corresponding
to the self-overlap of a pure state, i.e., $q_{\cal J}^{\alpha\alpha}$
in Eq.~(\ref{eq:qab})) remained zero.  Such a phase might
have, e.g., single-site magnetizations equalling zero
at low temperatures but two-spin correlation functions decaying
as a power law at large distances.

More likely, however, is that spin-flip symmetry {\it is\/}
broken for $d>d_0$ and $T<T_c(d)$ \cite{BY}.   In that
case the simplest scenario for the low-temperature
spin glass phase is the Fisher-Huse scaling/droplet 
picture \cite{FH86,HF87a,HF87b,FH88} (see also \cite{Mac,BM}),
in which a single pair of pure states is present.  In that
case, with periodic b.c.'s, CSD is
absent, and the metastate is concentrated on a single
mixed thermodynamic state, with each of the two pure states
having weight $1/2$.  This picture seems internally consistent.

We now consider possible many-state pictures.  In 
the standard SK picture, there is an overlap distribution
$P_{\cal J}(q)$ that exhibits non-self-averaging (NSA)
even after the thermodynamic
limit has been taken \cite{MPSTV1,MPSTV2,ultrametricity}; 
that is, it fluctuates with ${\cal J}$
even though it is a thermodynamic quantity.  Other features 
of this picture include
ultrametricity among {\it all\/} pure state
overlaps and a continuous part of
$P(q)$ (the average of $P_{\cal J}(q)$ over all ${\cal J}$) between 
$\pm q_{EA}$.  For details, see \cite{MPV}.

However, this standard SK picture {\it cannot} hold 
(in any dimension and at any temperature) \cite{NS96b} because the
translation invariance of $P_{\cal J}(q)$ combined 
with the translation ergodicity
of the underlying distribution of couplings implies that $P_{\cal J}(q)$ must
be self-averaged \cite{noteultra}.

This problem with the standard SK picture might sound like a
mere mathematical technicality -- for which one might hope
to find a technical solution. But in fact this picture has an
inherent conceptual flaw -- namely the basic
problem that a single state
$\rho_{\cal J}$ 
is simply not a rich enough description of the
$L\to\infty$ behavior of a thermodynamic
system where CSD occurs.  In such a picture, one
is in effect replacing with a single state
all of the information contained
in an entire distribution of states, i.e., the metastate.  We now
consider two nonstandard pictures, each of which arises
naturally in the context of the metastate approach
and the possible presence of CSD.

The first of these resembles the Fisher-Huse
picture in finite volumes, but has a very
different thermodynamic structure.  It is
a many-state picture, but unlike
in the mean-field picture each large volume (with
periodic boundary conditions) ``sees'' essentially only
one pair of states at a time (in Section~\ref{sec:pure} 
we discuss what it means for a finite volume
to ``see'' a thermodynamic state, pure or mixed).  In other words, for large
$L$, one finds that
\begin{equation}
\label{eq:possfive}
\rho_{\cal J}^{(L)}\approx {1\over 2}\rho_{\cal J}^{\alpha_L}+{1\over 2}\rho_{\cal J}^{-\alpha_L}
\end{equation}
where $-\alpha$ refers to the global spin-flip of pure state $\alpha$.
Here, the pure state pair (of the infinitely many present) 
appearing in finite
volume depends chaotically on $L$.  Unlike the droplet/scaling
picture, this new possibility exhibits CSD with periodic b.c.'s. 
In this ``chaotic pairs'' picture the (periodic b.c.) metastate is dispersed over (infinitely) many
$\Gamma$'s, of the form $\Gamma =
\Gamma^\alpha={1\over 2}\rho_{\cal J}^\alpha+{1\over 2}\rho_{\cal J}^{-\alpha}$.
The overlap distribution for each $\Gamma$ is the same:
$P_\Gamma = {1\over 2}\delta(q-q_{EA})+{1\over 2}\delta(q+q_{EA})$.
Like the Fisher-Huse picture, this scenario also seems internally consistent.
It is interesting to note that a highly disordered spin glass model 
\cite{NS94,NS96a} (see also \cite{CMB94}) appears to display just this
behavior in its ground state structure in sufficiently high dimension.

The last picture we discuss is a nonstandard SK-like
picture that resembles the standard SK picture
in finite volumes, but has an altogether different
thermodynamic structure.  This picture, which also assumes
infinitely many pure states, organizes them such that each 
$\Gamma=\sum_\alpha W_{\cal J}^\alpha\rho_{\cal J}^\alpha$.
The metastate $\kappa_{\cal J}$ is dispersed over many
such $\Gamma$'s, so that different $\Gamma$'s again appear
in different volumes, leading to CSD.  Unlike
the chaotic pairs picture, each $P_\Gamma$ 
depends on $\Gamma$ (because each $\Gamma$ is
now itself a nontrivial mixture of infinitely many
pure states).  However, the ensemble of
$P_\Gamma$'s (like the single $P_{\cal J}$ of the standard
SK picture) does {\it not\/} depend on
${\cal J}$ (again because of
translation invariance/ergodicity).
So the conventional meaning of NSA --- thermodynamic
quantities such as the overlap distribution depending
on ${\cal J}$ --- is replaced by a new notion:
not dependence on ${\cal J}$ but rather dependence on the
state $\Gamma$ within the metastate for fixed ${\cal J}$.  Moreover,
ultrametricity of overlaps among pure states
may be present within individual $\Gamma$'s,
but not for all of the pure states taken together.
A more detailed description of this nonstandard SK picture is given in 
Refs.~\cite{NS96c,NS97,NSBerlin}.

Given the results of \cite{NS96b}, the nonstandard SK picture is the only
remaining viable mean-field-like picture.  We have presented 
preliminary arguments (based on the invariance
of the ensemble of $P_\Gamma$'s 
with respect to ${\cal J}$; we refer the reader
to Ref.~\cite{NS97} for details) that already cast
some doubt on its validity, by demonstrating 
that the nonstandard SK picture requires
an enormous number of constraints to be simultaneously satisfied.  
In the next section we present
further arguments that more strongly rule it out 
as a viable possibility.

\section{Invariance of the Metastate}
\label{sec:invariance}

The main result of this section is a theorem on the
invariance of the metastate $\kappa_{\cal J}$ with respect to boundary
conditions that are flip-related.  Two (sequences of) 
b.c.'s are flip-related if,
for each finite $L$, there is some subset
of the boundary $\partial\Lambda_L$ whose flip transforms
one b.c.~for that $L$ into the other.  An obvious example
of flip-related boundary conditions are periodic and
antiperiodic; a second example is
any two fixed boundary conditions, i.e., where each
spin on the boundary is specified.  On the other hand,
periodic and fixed b.c.'s are not flip-related.

In the following theorem we continue to assume 
that the common distribution of the couplings $J_{xy}$ is
symmetric about zero, i.e., that $J_{xy}$ has the same distribution
as $-J_{xy}$, and that the external field is zero.

{\it Theorem.\/}  Consider two metastates constructed (at fixed, arbitrary
dimension and temperature, and using 
either the histogram method or the Aizenman-Wehr method)
using two different boundary conditions, with neither depending on ${\cal
J}$, on an infinite ($L_N\to\infty$) sequence of cubes $\Lambda_{L_N}$.
If the two different sequences of boundary conditions are 
flip-related, then the two metastates
are the same (with probability one --- i.e., for almost every ${\cal J}$).

\medskip

{\it Proof.\/}  We use the fact, discussed above, that along 
some ${\cal J}$-independent subsequence of volumes both
the histogram construction of metastates and the Aizenman-Wehr
construction have a limit, and that limit is the same.  Because the
Aizenman-Wehr construction averages over couplings
``at infinity'' (for details, see 
Refs.~\cite{NS96c,NS97,NSBerlin,Zurich,AW}), it rigorously
follows (using gauge transformation arguments like those used in the proof
of Theorem 3 in Ref.~\cite{NS92}) that the two metastates
must be the same.

\medskip

This is a striking result (despite the brevity of the proof),
with important physical consequences.  
It says, for example, that the periodic b.c.~metastate $\kappa_{\cal J}$ must
be the same as the {\it antiperiodic\/} b.c.~metastate.
In fact, if one were to choose (independently of ${\cal J}$) two 
{\it arbitrary\/} sequences of periodic and antiperiodic
b.c.'s, the metastates (with probability one) 
would {\it still\/} be identical.  In other words,
the metastate (and corresponding overlap distributions
constructed from it) at fixed temperature and dimension
is highly {\it insensitive\/} to boundary conditions.  

To appreciate the implications of this, consider the histogram
construction of the metastate.  The invariance of
the metastate with respect to different sequences
of periodic and antiperiodic b.c.'s means
that the frequency of appearance (in finite volumes) of various
thermodynamic states is (with probability one) {\it independent\/}
of the choice of boundary conditions.  Moreover, this same invariance property
holds (with probability one) among any two 
sequences of {\it fixed\/} boundary conditions
(and the fixed boundary condition of choice may even be allowed to
vary arbitrarily along any single sequence of volumes)!  
It follows that, with respect to changes of boundary conditions, the metastate 
is highly robust.

Of course, the insensitivity of the metastate with respect to 
changes of boundary conditions
would be unsurprising if there were only a single thermodynamic
state (e.g., paramagnetic) or a single pair of flip-related states
as in the droplet picture.  But it is difficult to see how our
result can be reconciled with the presence of {\it many\/} thermodynamic
states; indeed, at first glance it would appear to rule them out.

Nevertheless, we argue below that our theorem does {\it not\/} rule out the existence
of many states, but clearly puts severe constraints on the form of
the metastate (and overlap distribution function, which also
possesses this invariance property).  Our heuristic conclusion is that,
in light of this strong invariance property, any metastate 
constructed via coupling-independent b.c's can
support only a very simple structure.  As a consequence, we will argue
that this theorem effectively rules out the nonstandard SK picture.

To see that an {\it uncountable\/} set of pure states is
not ruled out (we will discuss countably infinite sets below),
consider the highly disordered ground state model \cite{NS94}
in high dimensions, which is believed to exhibit a version of the chaotic
pairs picture with uncountably many states.  Our invariance theorem 
applies to this model also, and so (e.g.) the   
periodic and antiperiodic metastates must 
be the same, even though we might {\it a priori\/} expect them
to be different.  By what mechanism could this happen?
The most natural possibility is that both the periodic
and antiperiodic b.c.~metastates are the same as the 
free b.c.~metastate \cite{notefree1} 
in which all relative signs between
the different trees in the invasion forest (see Refs.~\cite{NS94,NS96a}
for details) are equally likely. That is, each of these metastates 
consists of a {\it uniform distribution\/} on the
ground state pairs. Given that, it doesn't seem unreasonable that all
sorts of different b.c.'s should give rise to a similar uniform
distribution. Indeed, any fixed b.c.~{\it does\/} give a uniform distribution
on all {\it single\/} ground states \cite{NS94,NS96a}.

But this line of reasoning does appear to rule 
out the chaotic pairs picture with a
{\it countable\/} infinity of states.
In that case, of course, one can't have a uniform distribution 
(i.e., all equal, positive weights within the metastate).
So now suppose that for some ${\cal J}$ the periodic b.c.~metastate
assigns, for example, probability .39 to one pair of pure states, .28 to another, and so on.  
In other words, with periodic b.c.'s 39\% of the finite
cubes prefer
pair number 1, 28\% prefer pair number 2, etc.  So pair number 1 is the overall
``winner'' (among different finite volumes) in the periodic b.c.~popularity
vote.  

It now seems clear heuristically, though, that the popularity vote
by {\it antiperiodic b.c.'s\/} should come out differently;
it is unreasonable to suppose that pair number 1 be preferred by
39\% of the periodic b.c.~cubes and at the same time by 39\% of the 
antiperiodic b.c.~cubes!  The uniform distribution conclusion 
seems even more inevitable when
one considers that analogous arguments also apply to pairs
of arbitrarily chosen sequences of {\it fixed\/} boundary conditions.

We conclude that consistency between our invariance theorem
and the existence of (uncountably) many states
requires, in some sense, an equal
likelihood of the appearance (in the metastate) of all states, i.e., some sort
of uniform distribution on them.  Let us examine this
further.  We've already noted that different sequences
of volumes with fixed b.c.'s --- i.e., all
volumes having plus boundary conditions, all volumes having
plus on some boundary faces and minus on others, all
volumes with each boundary spin chosen by the flip of a fair
coin, and so on --- result in the same metastate.  
We note for future reference that the
term ``chaotic pairs'', which was chosen in reference
to spin-symmetric b.c.'s (such as periodic) should
be replaced here by ``chaotic pure states''; 
i.e., in this picture,
the Gibbs state in a typical large volume $\Lambda_L$ with fixed b.c.'s will 
be (approximately) a single pure state that
varies chaotically with $L$.  But we expect that the mixed
state $\rho_{\cal J}$, which is the {\it average\/} over
the metastate \cite{NS96b,NS96c,NS97,NSBerlin}
\begin{equation}
\label{eq:rhoav}
\rho_{\cal J}(\sigma)=\int\ \Gamma(\sigma)\kappa_{\cal J}(\Gamma)\ d\Gamma\quad ,
\end{equation}
would be the same for periodic and fixed b.c.'s.  One can also think
of this $\rho_{\cal J}$ as the average thermodynamic state,
$N^{-1}(\rho_{\cal J}^{(L_1)}+
\rho_{\cal J}^{(L_2)}+\ldots,\rho_{\cal J}^{(L_N)})$, in the limit
$N\to\infty$.

Now consider the mixed boundary condition in which
{\it every\/} fixed b.c.~on the boundary of each
$\Lambda_L$ is given equal weight.  If there
are (uncountably) many pure states  
present, then in a typical large volume one would expect to
see a Gibbs state which approximates a continuous 
mixture over the pure states
(cf.~Possibility 3 or 4 discussed in Ref.~\cite{NS96c}).
But we still expect that the average over the mixed 
b.c.~metastate  would be the same $\rho_{\cal J}$ 
as for the fixed b.c.~metastate, the
periodic b.c.~metastate, and so on.  
That is, the average over the metastate should be
even more robust than the metastate itself, i.e., it should
be the same for metastates constructed through {\it any\/}
two sequences of (coupling-independent) b.c.'s, not just
flip-related ones.

Although logically possible, it seems
unreasonable that this last (mixed b.c.~with
all fixed b.c.'s given equal weight) metastate, chosen
from a maximally uniform mixture of boundary conditions, 
can have anything other than
a uniform distribution over the pure states.
But, as just pointed out, this distribution 
should be the same for this as for all the other
metastates under discussion.  (We caution the
reader that, unlike the case of the strongly
disordered model \cite{uniformnote}, we do not have a precise
sense in which this distribution can be
defined to be uniform.  For that reason, this part of the argument 
must be regarded as heuristic.)

With these points in mind, we now turn to a discussion 
of the nonstandard SK picture, and other possible
mixed state scenarios.

The nonstandard SK picture requires 
(cf.~Eq.~(\ref{eq:sum})) that the $\Gamma$'s 
appearing in the metastate be of the form 
$\sum_\alpha
W_{\cal J}^\alpha\rho_{\cal J}^\alpha$, with at least
some subset of the weights $W_{\cal J}^\alpha$ in each 
$\Gamma$ nonzero and unequal.  We would then have a 
situation like the following.  With periodic b.c.'s, say, the fraction
of $L_j$'s for which the finite volume Gibbs 
state in $\Lambda_{L_j}$ puts (e.g.) at least 84\%
of its weight in one pair of pure states (but with that pair not
specified) is 0.39.  But then it must also be the case that with
antiperiodic b.c.'s the fraction of volumes for which the finite
volume Gibbs state puts at least 84\% of its weight in some unspecified
pair is still exactly 0.39!  Moreover, the same argument must
apply to any ``cut'' one might care to make; i.e., one constructs
the periodic b.c.~metastate and finds that $x\%$ of all finite volumes 
have put $y\%$ of their weight in $z$ states, with $z$ depending
on the (arbitrary) choice of $x$ and $y$.  Then this must be true also
for all volumes with antiperiodic b.c's; and similarly (but possibly separately)
among all pairs of fixed b.c.~states.

Once again, the only sensible way in which this could happen
would be for the selection of states to be relatively insensitive (in
some global sense) to the choice of boundary conditions, i.e., 
for the b.c.'s to choose the states
in some ``democratic'' fashion without favoritism so that
$\rho_{\cal J}$, the average over the metastate, should be
some sort of uniform mixture of the pure states, as before.  
However, unlike in the chaotic pairs picture discussed earlier, 
we claim that this {\it cannot\/} happen when the $\Gamma$'s
are (nontrivial) mixed states.  

The reason for this is that the metastate has
a strong covariance property \cite{AW} (see also \cite{NS97}) in
which the $\Gamma$'s must transform in a 
specified way under an arbitrary finite change in 
the coupling realization.  Under this finite
change, the ensemble $\kappa_{\cal J}(\Gamma)$
transforms (as would any probability measure)
according to the change of variables $\Gamma \to \Gamma'$.  Here, $\Gamma'$ 
is the thermodynamic state with correlations
$\langle \sigma_A \rangle_{\Gamma'} = 
\langle \sigma_A e^{-\beta \Delta H} 
\rangle_\Gamma / \langle e^{-\beta \Delta H} \rangle_\Gamma$,
where $\Delta H$ is the change in the Hamiltonian.

Under this change of variables,
pure states remain pure and their overlaps
don't change.  However, the weights which appear
in each $\Gamma$ {\it will\/} in general change,
as one would expect.  To see this,
consider a particular $\Gamma$ having a discrete pure state 
decomposition
\begin{equation}
\label{eq:sumgamma}
\Gamma=\sum_\alpha W_{\Gamma}^{\alpha}\rho_{\cal J}^\alpha (\sigma)\ ,
\end{equation}
with many nonzero
weights $W_\Gamma^\alpha$.
Suppose that one chooses a particular coupling $J_{xy}$
and imposes the transformation $J_{xy}\to J'_{xy}=J_{xy}+\Delta J$.
Then the weight $W^\alpha$ (within $\Gamma$) of the pure state $\alpha$
will transform for each $\alpha$ as
\begin{equation}
\label{eq:walpha}
W^\alpha\to W'^\alpha=r_\alpha W^\alpha/\sum_\gamma r_\gamma W^\gamma
\end{equation}
where 
\begin{equation}
\label{eq:ralpha}
r_\alpha=\left\langle\exp(\beta\Delta J\sigma_x\sigma_y)\right\rangle_\alpha=
\cosh(\beta\Delta J) +
\langle\sigma_x\sigma_y\rangle_\alpha\sinh(\beta\Delta J)\ .
\end{equation}

In either the droplet/scaling or the chaotic pairs picture,
there are in each $\Gamma$ only two pure states (depending on $\Gamma$ in
chaotic pairs), each with weight $1/2$.  Because all even correlations
are the same in each pair of (flip-related) pure states, the transformation of
Eq.~(\ref{eq:walpha}) leaves the weights unchanged.  

However, in nonstandard SK there exist pure states
within each (mixed) $\Gamma$ with relative domain walls, so
that they differ in at least some even correlation functions.
But this then rules out that $\rho_{\cal J}$ 
must always be a uniform mixture of the pure
states, because a suitable change of couplings
will shift the weights for each $\Gamma$ in such a way
that the distribution over pure states of $\rho_{\cal J}$ also shifts.
(This reasoning can be made rigorous, but because
other parts of the argument are heuristic,
we omit a proof.)  

In other words, we argued above
that the invariance of the metastate with respect to boundary conditions left
open, as the only reasonable possibility for the presence of
many pure states,  that $\rho_{\cal J}$, the average over the
metastate, be some sort of uniform mixture over
the pure states.  This must be true for any ${\cal J}$
(with probability one), so the weight distribution over
all pure states must also be invariant with respect to changes in
${\cal J}$.  But this invariance is inconsistent with the transformation
properties of the $\Gamma$'s with respect to finite changes
in ${\cal J}$: if there are multiple pure states in 
the $\Gamma$'s, with the pure states
in each $\Gamma$ not having the same even correlations
(i.e., they have relative domain walls), then their relative weights
must vary (as expected) with changes in the coupling realization.
This leads to a contradiction, and therefore rules out not only
nonstandard SK but any picture in which the $\Gamma$'s
are a nontrivial mixture of pure states.

%
%
Our conclusion, based on the above combination of both
rigorous results and heuristic arguments, is that
the nonstandard SK picture cannot be valid in any
dimension and at any temperature.  More generally,
the many invariances of the spin glass metastate
cannot support {\it any\/} picture in which thermodynamic
mixed states (other than a single flip-related pair)
are seen in finite volumes.  

Given that the only reasonable possibilities remaining
(that display broken spin flip symmetry) are the droplet/scaling
picture and the chaotic pairs picture, we conclude that
the overlap distribution function $P_{\Gamma}$ 
\begin{equation}
\label{eq:PJgamma}
P_{\Gamma}(q)=\sum_{\alpha,\gamma}W_{\Gamma}^\alpha W_{\Gamma}^\gamma
\delta(q-q^{\alpha\gamma})\quad 
\end{equation}
can at most be a pair of $\delta$-functions at $\pm q_{EA}$ for each
$\Gamma$; i.e., in each finite volume the overlap between
pure states that appear in that volume is just that pair
of $\delta$-functions.  This will be the case regardless
of whether there is only a single pair or 
uncountably many pairs of pure states.
We will discuss this further in Section~\ref{sec:pure},
but first we turn to another topic.

In the next section we present
a simple scaling approach that provides both a plausibility argument
and also a physical starting point for
understanding the ``chaotic pairs'' many-state picture introduced in
Refs.~\cite{NS96c,NS97,NSBerlin}.  It is 
important to note that this scaling picture 
is consistent with the Fisher-Huse droplet picture \cite{FH86,FH88} 
for appropriate values of the new scaling exponents, but for other
values can give rise to a different thermodynamic picture.

\section{A Scaling Approach to the Chaotic Pairs Picture}
\label{sec:scaling}

We have argued above that with periodic boundary conditions, one should
see at most a single pair of flip-related pure states in a large 
volume.  As already discussed, this leaves open the possibilities
of either a single pure state (e.g., but not necessarily, 
a paramagnet), a single
pair of pure states (as in the droplet picture), or the
chaotic pairs many-state picture discussed above.  We now
present a simple extension of earlier scaling/droplet
arguments \cite{FH86,FH88} which is consistent with this last possibility,
and also provides a possible scenario for the spatial structure of
domain wall configurations among the 
ground states.

The object here is to obtain estimates on the difference
in energy or free energy between 
the lowest-lying state in a fixed volume and
the next higher one.  The appearance at 
nonzero temperature of multiple (non-spin-flip
related) states in a single (large) volume requires that
the energies of the lowest-lying states differ by order one.
If, on the other hand, the ``minimal'' energy difference scales as
some positive power of the system size, then one will see
at most a single pair of states in any given box
(with spin-symmetric boundary conditions,
such as periodic).

To analyze the appearance in finite volumes, and at very low
temperature, of infinite volume pure states,
as in Eq.~(\ref{eq:sum}), we will consider infinite volume
{\it ground\/} states restricted to the cube of size $L$, 
with a fixed boundary condition $\hat\sigma$ chosen
independently of the couplings.  In our analysis below
we will treat the boundary spins as chosen randomly and independently
of the couplings --- but for a nonrandom fixed b.c.~such as plus,
the same arguments go through
with minor modifications.

Although there may a priori be 
infinitely many infinite volume ground states,
the number of distinct restrictions to the cube is finite
and its logarithm should be of order $L^{d-1-\phi}$
for some $\phi$. The scaling exponent 
$\phi$ (with $0\le\phi\le d-1$)
may be understood in another way: the minimum number of
spins on the surface of the cube that differ between two infinite
volume ground states, whose spins disagree at (or near)
the origin \cite{noteground}, should scale as $L^\phi$.
These two states should correspondingly
differ in the bulk by a number of spins of (at least) 
order $L^{\phi+1}$.

If there exists only a single pair of flip-related
ground states (as argued in Refs.~\cite{HF87a,HF87b}), then
$\phi=d-1$.  In the highly disordered spin glass
model of Refs.~\cite{NS94,NS96a} (see also Ref.~\cite{CMB94}), 
it appears that $\phi=d-1$ below eight dimensions
while $\phi=3$ above eight dimensions.

Let us examine the exponent $\phi$ more closely.  
Although {\it a priori\/} there seems to be no reason to
exclude the possibility that $\phi=0$, there
are several arguments indicating otherwise.
(Note also that $\phi=0$ would saturate the possible growth rate
of the number of distinguishable ground states in any finite volume since
the logarithm of this number cannot exceed order $L^{d-1}$.)
If $\phi=0$, then spins living in regions between domain walls
would exist in one-dimensional tube-like objects.  It
seems very unlikely that such tubes could be stable; i.e.,
eventually such a tube should encounter a fluctuation which 
destroys its structure.  A second and somewhat different argument
uses the fact that $\phi$ should be bounded
from below by the exponent $\theta$ introduced by
Fisher and Huse \cite{FH86,FH88}, which governs the minimal interface
free energy between different pure states on a length scale
$L$; i.e., this minimal free energy is presumed to grow as $L^\theta$.
It is not difficult to see, then, that 
$\phi\ge\theta$.  However, it was argued in Refs.~\cite{FH86,FH88} 
that the inequality $\theta>0$ is necessary in 
order for spin flip symmetry to be broken at positive temperature.
In what follows we therefore always assume that $\phi>0$.

Before considering the EA model itself, we first treat the
much simpler case of a homogeneous Ising ferromagnet with
fixed b.c.'s chosen at random.  First we consider the energy difference
between the plus and minus ground states (with interface
ground states temporarily not considered).  Here there is
no bulk energy difference, and $\phi=d-1$.
Because of the randomness of the b.c., the boundary energy
difference is of order $L^{\phi/2}$.  The conclusion in this
case \cite{En90} (see also \cite{NSBerlin}) 
is that the total energy difference is
also $L^{\phi/2}$ and thus with random b.c.'s one does
not see a mixture of the plus and minus states but only one of
them (chosen by the sign of the boundary energy) chaotically changing
with $L$.  

What about seeing interface states?  Here, the appropriate
bulk energy difference between the constant ground states and the interface
states scales as $L^{d-1}$ (with $\phi$ 
the same as before) and so the bulk energy difference
dominates the boundary energy difference.  In this case the
total energy difference between the homogeneous state and the
lowest-lying interface state is of order $L^{d-1}$.  As a result,
all interface states are ``invisible'' in the random b.c.~finite volume
ferromagnet \cite{NSBerlin,En90}.

We now consider the EA Ising spin glass from this point of view.
That is, we consider the energies of the restrictions of all
infinite volume ground states to the $L^d$ cube centered
at the origin.  As before, we divide the energy into a bulk
and a boundary part, and ask how the energy difference between
the lowest-energy and next-lowest-energy state scales with $L$.  Consider
the state $\rho^*$ with minimum total energy (subject to the
fixed boundary condition) and the state
of next lowest energy that differs from $\rho^*$
near the origin.   By the definition of $\phi$, the two states
differ by at least $L^{\phi+1}$ spins in the bulk and
by $L^\phi$ spins on the boundary.

To estimate the energy differences between low lying states,
we will separately consider the boundary energy coming from the
couplings between $\hat\sigma$ and the adjacent spins in the
cube, and the bulk energy difference (from the remainder of the
finite volume Hamiltonian).   If there were no bulk energies
to consider, then one might expect that two states which
differ by $L^\phi$ spins on the boundary would
typically differ by an overall energy of order $L^{\phi/2}$.  
If this were indeed the case for the two lowest-lying states
in almost any volume, then one would see only one
state per volume (for fixed boundary conditions).
However, since one is doing a minimization
problem which includes bulk energies as well, it is
not at all obvious {\it a priori\/}  that this will
happen.  In particular, there might be some
delicate cancellation between bulk and boundary energies.  

We will now, however, present a specific scenario
in which an explicit calculation shows that 
the lowest-lying states, in a volume with
fixed boundary conditions chosen 
independently of the couplings, 
do indeed have an energy difference 
of order $L^{\phi/2}$.  This example
is presented as a plausibility argument and demonstrates
one way in which this can occur, but is not meant to
imply that it can occur in {\it only\/} this way.

Consider then a scenario in which the spin at the origin
belongs to a cluster, not intersected by {\it any\/}
domain walls, whose intersection with the boundary as before is of size
$L^\phi$.  We denote that cluster ${\cal S}_0$.  Suppose further
that $\rho^\alpha$ is a general infinite volume ground state, and that
$E_L({\alpha})$ is the energy --- including both the boundary and
bulk components --- of $\rho_\alpha$ restricted to $\Lambda_L$,
the $L^d$ cube centered at the origin.

The energy $E_L({\alpha})$ can therefore be written
\begin{equation}
\label{eq:el}
E_L({\alpha})=-\sum_{x\in{\cal S}_0\cap\partial\Lambda_L}\sigma_x^{\alpha}{\overline\sigma_x}
- \sum_{x\in\partial\Lambda_L\backslash{\cal S}_0}\sigma_x^{\alpha}{\overline\sigma_x} +E_L^b(\alpha)\quad ,
\end{equation}
where the first term is the contribution from the spins in the cluster
${\cal S}_0$ on the boundary $\partial\Lambda_L$, 
the second term is the surface energy 
contribution from all other boundary spins, and the final term 
is the energy contribution of the bulk spins.
More precisely, $\partial\Lambda_L$ is the set of sites $x$ inside
$\Lambda_L$ with a nearest neighbor $y$ outside $\partial\Lambda_L$
and ${\overline\sigma_x}$ is the boundary spin ${\hat\sigma_x}$ 
times $J_{xy}$.
Eq.~(\ref{eq:el}) can be rewritten as
\begin{equation}
\label{eq:twoterms}
E_L({\alpha})=-\eta(\alpha)Z_L\sqrt{|{\cal S}_0\cap\partial\Lambda_L|} +Y_L(\alpha)
\end{equation}
where three new variables have 
been introduced:  $\eta(\alpha)=\pm 1$ represents the
sign of the spin at the origin in ground state $\alpha$, $Z_L$ is 
(approximately) a Gaussian random variable with 
zero mean and unit variance, and $Y_L(\alpha)$ depends
both on the bulk energy of $\alpha$ and on the rest of 
the boundary spins (i.e., those not
included in the first term).  

In going from Eq.~(\ref{eq:el}) 
to (\ref{eq:twoterms}) we used the fact that the boundary
condition consists of fixed random spins, chosen independently of $\alpha$.  
The crucial observation is that the random variables $Z_L$, which arise from
the random boundary conditions, are independent of the spectrum of
the (mostly) bulk energies $Y_L(\alpha)$.  We now show that, regardless of the
number and distribution of the $Y_L(\alpha)$'s as $\alpha$ varies, there 
will be no strong cancellations between the two terms
(with probability close to one).

Consider the ground state whose energy in Eq.~(\ref{eq:el}) is the minimum,
and also the ground state which 
has the next higher energy, and is {\it required\/} 
to have a relative spin flip with 
respect to the lowest energy state at the origin.
We then have
\begin{equation}
\label{eq:diff}
\left|\min_{\gamma:\sigma_0^\gamma=-1}E_L(\gamma)-\min_{\alpha:\sigma_0^\alpha=+1}E_L(\alpha)\right|
=\left|2Z_L\sqrt{|{\cal S}_0\cap\partial\Lambda_L|}
+Y_L^--Y_L^+\right|\quad ,
\end{equation}
where $Y_L^-$ and $Y_L^+$ are the bulk 
plus remainder boundary energies of the two lowest-lying
states with a relative spin flip at the origin.

Since $Z_L$ and $Y_L^--Y_L^+$ are 
functions of {\it disjoint\/} sets of the random boundary
spins, they are independent random variables. Hence, 
variances add and the effect of 
$Y_L^--Y_L^+$ on the random variable 
$2Z_L\sqrt{|{\cal S}_0\cap\partial\Lambda_L|}$
can only be to {\it increase\/} the spread of its distribution. 
This allows us to conclude that with probability close to one
(i.e., for most choices of the 
boundary spins) the expression on the right-hand side
of Eq.~(\ref{eq:diff}) is of order (at least) 
$\sqrt{|{\cal S}_0\cap\partial\Lambda_L|}$, i.e., of
order $L^{\phi/2}$. As long as $\phi > 0$, which is part of our scenario, this growth with
$L$ in the spacing of the low-lying spectrum of ground states argues for the appearance
at small positive temperature of only a single pure state in large finite volume Gibbs
states with fixed b.c.'s (that are independent of the couplings).

\smallskip

The above argument is instructive in several respects.  It demonstrates
that, given the condition that no domain wall separates the origin from
the boundary of the box, there can be no miraculous ``conspiracy'' under
which bulk and boundary energies cancel out to order one.  It does require
a strong condition, namely that all domain walls, in the union of all
symmetric differences over all ground states, do not form any closed
and bounded regions.  As stated above, this is a {\it sufficient\/} condition
for the scaling argument given above to work, but we see no reason
at this point why it should be a {\it necessary\/} condition in order 
for the conclusions to be valid.   

Nevertheless, it provides one interesting scenario for the spatial structure of ground states
and domain walls if many states should exist.  Interestingly, in the only
example of which we're aware in which a finite dimensional spin
glass apparently {\it does\/} 
possess many states in high dimensions --- the highly
disordered ground state model of 
Refs.~\cite{NS94,NS96a} --- exactly this structure
occurs!  These considerations provide a possibly fruitful avenue
for future investigations.

\section{Pure States in Finite Volumes:  What's Going on Here?}
\label{sec:pure}

In this section we address what it actually means, in an operational
sense, to ``see'' a pure state --- which formally is an infinite volume
object --- inside a finite volume.  We then use that analysis to answer
a glaring question:  if states and overlaps in finite volumes
are restricted to, at most, a single pair of flip-related pure states
and a pair of $\delta$-functions at $\pm q_{EA}$, respectively, then
what are the many numerical simulations (e.g., \cite{Parisi3,MPR94,MPRR,MPR97,RBY,Caracciolo})
and experiments (\cite{Orbach1,Orbach2}) that appear to see a more
complicated state and overlap structure actually seeing?

Our main point will be that pure state structure can and does manifest
itself in finite volumes, and governs the 
physics at finite length scales.  
Conversely, observations made in large, finite volumes
must in turn reveal the thermodynamic
structure and the nature of ordering of the system --- if sufficient care is given to
the analysis of those observations.   Indeed, were both the
above statements not true, it would be difficult to see why
the study of thermodynamics would be of any interest to physics.

While the above assertions have long been noncontroversial for
most statistical mechanical systems and models, there remains
considerable confusion in the case of spin glasses \cite{noteMPR}.
At least part of the problem is that reliance on 
the overlap structure alone can at best give only partial --- and
sometimes misleading --- information
on the thermodynamics of realistic spin glass models
\cite{NS96c,NS97,HF87a,HF87b}.
A second problem is that, as we have emphasized
in previous papers \cite{NS96c,NS97}, the connection
between finite- and infinite-volume behavior may be
more complex and subtle in spin glasses than in
simpler systems. An analysis of this connection thus
deserves more thought than a simple attempt to 
sever the link altogether between the
two behaviors (as in Appendix I of \cite{MPR97}).  
So in this section we will expand on previous 
discussions \cite{NS97} to further clarify these issues.

A thermal state, whether pure or mixed, is completely
specified by the set of all of its (1-point,
2-point, 3-point, $\ldots$) correlation functions.
In a finite volume, a state will
manifest itself through the appearance of a particular
set of such correlations.  Because boundary effects will
invariably alter or distort (compared to an infinite-volume
state) these correlations in some region 
(whose size will depend on the specifics
of the Hamiltonian, temperature, dimension, etc.), one must
always be careful to examine them in a volume
small enough so that these ``distortion'' effects are 
negligible.  In other words, 
the boundary should be sufficiently far from
the region under examination so that an accurate picture
of the thermodynamics can be obtained \cite{notethermo}.

So, for example, even in the paramagnetic state, one
would measure nonzero magnetizations at interior sites
in the vicinity of a boundary on which all spins are
fixed (e.g., to be $+1$).  As the boundary moves farther out,
subsequent measurements at those same sites would
find their magnetization tending to zero.

It is not unusual, even for comparatively simple systems,
for boundary effects to penetrate more deeply into the interior
than a shallow ``boundary layer''.  
Consider the example of the two-dimensional 
uniform Ising ferromagnet.  
It is known \cite{A80,H81} that
this system has only two pure states --- the translationally
invariant positive and negative magnetization states ---
for all $0<T<T_c$.  Suppose now that on a square of side $L$ one were
to impose fixed boundary conditions such that all spins
on the right half of the boundary are $+1$ and all
spins on the left are $-1$.  This will
impose a domain wall on the system, whose maximum (and typical) deviation
(from the vertical line passing through the origin) will
scale as $L^{1/2}$  (see Figure 1).  So for all large $L$ the system gives
the {\it appearance\/} of having a pure state with a domain 
wall \cite{notedomain}; indeed, the domain wall
always stays quite far from the (vertical) boundaries.
However, if one were to look at {\it any\/} fixed, finite region,
then as the size $L$ of the square grows, the domain wall
eventually moves outside the fixed region, and one would see
only a mixture of the positive and negative translationally
invariant states. The (equal, in this case, as $L\to\infty$)
weights in the mixture correspond to the probabilities
of the domain wall thermally fluctuating to the left or to the
right of the fixed region.

So in this example the domain wall is an artifact of the imposed boundary
condition, and has nothing to do with any thermodynamic structure
or low-temperature ordering properties of the system.  Moreover,
consideration of the spin configurations 
over the entire square
would lead to incorrect conclusions about the pure
state structure.  This illustrates
our contention that {\it in order to arrive at an accurate picture
of the thermodynamic structure and the nature of ordering of a system,
one must focus attention on a fixed ``window'' near the origin
(which may be arbitrarily
large, but is small compared to the entire
volume under consideration).}

This conclusion is especially important when evaluating,
and drawing inferences from, overlap functions.  
A more detailed discussion is given in the Appendix
of Ref.~\cite{NS97}, to which we refer the reader;
here we will only reiterate an illuminating example
due to van~Enter \cite{Enpriv}, which in turn
extends an earlier example due to Huse and Fisher \cite{HF87a}.
Consider the overlap distribution of 
an Ising antiferromagnet in two dimensions with periodic boundary conditions.
For odd-sized squares the overlap is equivalent 
(by the obvious gauge transformation) to 
that of the ferromagnet with periodic boundary conditions, and for even-sized
squares it is equivalent to that of the
ferromagnet with antiperiodic boundary conditions.  If the
overlap distribution were computed in the {\it full\/} square,
it would therefore oscillate between two different
answers (one a sum of two $\delta$-functions at plus or minus
the square of $M^*$, the spontaneous magnetization, and the
other a continuous distribution between $\pm (M^*)^2$).  
On the other hand, computing overlaps in boxes
which are much smaller than the system size would give rise 
in this example to a well-defined answer --- i.e., the two 
$\delta$-function overlap
distribution --- which provides a more
accurate picture of the nature of ordering in this system. 

With these remarks in mind, we now turn to the finite-dimensional
Ising EA spin glass.  Essentially all the simulations
of which we are aware compute the overlap distribution in
the {\it entire\/} box.  Boundary conditions are chosen
independently of the couplings, and are usually periodic.
Given our conclusion that, under these circumstances, at most
a pair of flip-related pure states will appear in almost
any finite volume, we suspect that the overlaps computed
over the entire box are observing
domain wall effects arising solely from the imposed boundary conditions,
rather than revealing the actual spin glass ordering.
(This is the reason why in Section~\ref{sec:scaling}
we looked only at states with relative domain walls
in the vicinity of the origin.)

In other words, if overlap computations were measured in
``small'' windows far from any boundary, one should find only
a pair of $\delta$-functions.  One way to test this would be to fix a
region at the origin, and do successive overlap computations
in that fixed region for increasingly larger boxes 
with imposed periodic boundary conditions; as the
boundaries move farther away, the overlap distribution within
the fixed region should tend toward a pair of $\delta$-functions \cite{noteBinder}.

It is important to clear up one other misconception.
It was asserted at the end of Section~2 in
Ref.~\cite{MPR97} that ``after Ref.~\cite{RBY} 
one has to argue that the physics must change after some very
large length scale ...in order to claim that the mean field
limit is not a good starting point to study the realistic
case of finite dimensional models...''.  Although, of course,
this changeover may well occur, it is at least as likely that
it doesn't \cite{notefinitesize}, 
and that nontrivial overlaps will be seen for all large $L$
(as the uniform ferromagnet domain wall example illustrates).
The real problem is in some sense the opposite:  namely,
that overlap computations are not
being done in {\it small\/} enough regions
to provide an accurate picture of spin glass ordering.

\section{Conclusions}
\label{sec:conclusions}

In our previous papers \cite{NS96b,NS96c,NS97}, we 
showed that spin glasses may be more complex --- in the
relation between their behavior in finite and infinite volumes ---
than had previously been noted in the literature.  
In the present paper, we have presented arguments 
indicating that, in a different sense, spin
glasses are more {\it simple\/} than had previously
been claimed in much of the literature.

Our main conclusion is that, for realistic spin glass
models such as Ising Edwards-Anderson, any large finite
volume (with say spin-symmetric b.c.'s, such as periodic,
chosen independently of the couplings) will display at most a single
pair of flip-related pure states.  This may correspond to
either a single pair of pure states in total, as in the
droplet/scaling picture \cite{FH86,HF87b,FH88}, or to the ``chaotic
pairs'' picture introduced in Ref.~\cite{NS96c} and elaborated upon
in Refs.~\cite{NS97,NSBerlin}.

This rules out the nonstandard SK picture
also introduced in Ref.~\cite{NS96c} and elaborated upon
in Refs.~\cite{NS97,NSBerlin}.  Combined with our earlier
result \cite{NS96b} ruling out the standard SK picture,
we conclude that {\it the thermodynamic structure and the nature
of spin glass ordering, whether in finite or
infinite volumes, cannot be mean-field-like in any
dimension and at any temperature\/}.

The argument leading to this conclusion followed a
theorem, presented in Section~\ref{sec:invariance},
that the metastate for fixed ${\cal J}$ is invariant
with respect to arbitrary choices of flip-related boundary
conditions (such as periodic and antiperiodic).
It was then argued that only the simplest
pure state (and corresponding overlap) structures
could be so robust
\cite{notefree2}.
The only reasonable scenario under
which (uncountably) many states could then appear is 
that, statistically, the states
are insensitive to the boundary conditions.
That is, the 
metastates would be generated (as in the highly disordered 
ground state model) through some kind of random 
fair-coin-tossing process.

We argued in Section~\ref{sec:pure} that overlap
computations should be done in small interior
boxes (surrounded by much larger boxes where
the boundary conditions are actually imposed) in
order to remove boundary effects and
get a picture of spin glass ordering
that is not misleading.  We expect
that (with periodic b.c.'s) for those dimensions and temperatures
where $q_{EA} \neq 0$, this procedure would result in a single
pair of $\delta$-functions at $\pm q_{EA}$ \cite{noteconstruction}.

We also presented in Section~\ref{sec:scaling}
a scaling argument that shows
how a ``chaotic pairs'' (or chaotic pure states,
under fixed b.c.'s) picture can arise.  We provided an
explicit calculation that supported this picture
under the sufficient (but not necessary) condition
that the union of domain walls between {\it all\/} pairs of
pure states form no closed and bounded regions.
Interestingly, exactly such a structure is present in the
only example of a nontrivial short-ranged spin glass 
model known to have many 
ground states --- i.e., the highly disordered spin glass
model of Refs.~\cite{NS94,NS96a} (see also \cite{CMB94}).

Given that an overlap structure computed in an
entire finite volume (as opposed to that computed within
a smaller window) might be nontrivial due only
to boundary effects, it cannot yield definitive information
on the ordering of the spin glass phase. 
Furthermore, there is no a priori reason
to expect that it would display any
exotic or intricate properties such as ultrametricity, 
or in general bear any
particular resemblance to the mean field
structure observed for the SK model.  However, the
domain walls responsible for this overlap
structure (if present) could have an observable, although
perhaps nonuniversal, effect on dynamics.  We will
explore this issue in a future paper.

{\it Acknowledgments.\/}  This research was partially 
supported by NSF Grant DMS-95-00868 (CMN) 
and by DOE Grant DE-FG03-93ER25155 (DLS).

\begin{figure} \caption{A typical spin configuration in a $2d$ Ising ferromagnet 
at positive temperature below $T_c$, with fixed spin boundary
conditions that are $+1$ on the right half of the boundary
and $-1$ on the left half.  The maximum (and typical) deviation of the
induced domain wall from the vertical line through the origin
is $O(L^{1/2})$.  This domain wall persists on all length scales
but is unrelated to the low-temperature ordering.  It will miss
a sufficiently small ($o(L^{1/2})$) window about the origin; examination
of the order parameter inside only this window will correctly 
capture the thermodynamics.  (In particular, one can examine any
fixed finite region as the boundaries move far away.)
This sketch depicts a relatively small square; for large
$L$, the domain wall would be virtually indistinguishable
from a straight line through the origin (on the scale $L$
of the entire square), and the window would be extremely small 
(on that scale).}
\end{figure}


\begin{references}
\bibitem{subsequence}
It should be noted that currently existing proofs require
not only a subsequence $L_1,L_2,\dots,L_N,\dots$ of cube
sizes, but possibly also a subsequence of $N$'s when taking the
histogram limit.  However, the crucial point is that
this subsequence of cube sizes, even if necessary in some instances, 
remains independent of ${\cal J}$.
\bibitem{noteground}
This is clearly a subset of all pairs of the ground states 
that are distinct within
the cube $\Lambda_L$; the reason for this restriction will be
discussed in Section~\ref{sec:pure}.
\bibitem{noteMPR}
In Appendix~1 of Ref.~\cite{MPR97}, for example, 
it is argued that the pure state, 
or thermodynamic, structure 
is merely a mathematical infinite-volume construct that has little 
or no physical
relevance to real (finite-volume) systems such as spin glasses.  We believe 
those arguments to be misleading,
and indeed, misdirected in that the metastate approach 
precisely does connect the
behavior of observable quantities in finite volumes with the thermodynamic
structure of the system.  (Moreover, the suggestion in that same reference 
that the Boltzmann-Gibbs probability distribution 
does not even exist in the infinite-volume
limit for many disordered systems 
is simply incorrect.)  It is, for example, a misconception that the
behavior of correlation functions is more physical 
or less ``metaphorical'' (cf.~Appendix~1 
of Ref.~\cite{MPR97}) than thermodynamic states.  Indeed, the two are
simply different labels for the same object, in the same way that one
can talk either of the probability distribution of a random variable 
or the set of its moments.
\bibitem{notethermo}  
This should not be confused with the fact that,
if many pure states are present, then changes in boundary conditions
can change the state everywhere in the volume, including the region
about the origin.  In this situation, boundary conditions can select
the thermodynamic state in the interior; but in order to see which 
state has been selected, one must still measure correlations in
a region about the origin sufficiently far from the boundaries.
\bibitem{noteBinder}
Although all direct numerical computations of $P_{\cal J}(q)$
(and $P(q)$) of which we are aware compute
overlaps in the full volume, at least one computation has been reported
\cite{MPRR,MPR97} that does examine a type of overlap measure,
called the Binder cumulant, constructed on restricted subvolumes. 
Although strictly speaking the measurement reported has a dynamical
component, it may contain potentially interesting and currently unexplained
information on the equilibrium spin glass.  However, the limited
nature of the measurements done to date seem to us insufficient
grounds for ruling out the droplet/scaling picture, as asserted in 
\cite{MPRR,MPR97}.
\bibitem{notefinitesize}
The possibility that finite size effects might be persistent
in systems with quenched disorder was also noted in Ref.~\cite{MPR97}.
\bibitem{noteconstruction}
We discussed in the Appendix to Ref.~\cite{NS97} various subtleties
associated with the precise method of construction of the overlap
distribution.  In this paper we have referred only to the case
where the overlap is computed in finite volumes using
the replica measure $\rho_{\cal J}^{n(L)}$ discussed in
that paper.  If replica non-independence \cite{NS96c,NS97}
were present, as would be the case if the chaotic pairs
picture were to hold, then one could construct
a different infinite-volume overlap distribution by breaking
replica symmetry {\it after\/} the infinite-volume
limit is taken (cf.~construction~2 of Ref.~\cite{NS96b}).
This would be the replica overlap for the
average $\rho_{\cal J}$ of the metastate, and it
would be the same not only for almost all flip-related
boundary conditions but also, at the same time,
for almost every ${\cal J}$.  Given that, the only
reasonable possibilities for this overlap function
within the chaotic pairs scenario
would be either a single $\delta$-function at the origin, or
(less likely, we believe) a continuous distribution
between $\pm q_{EA}$ with {\it no\/} $\delta$-function spikes.
\end{references}
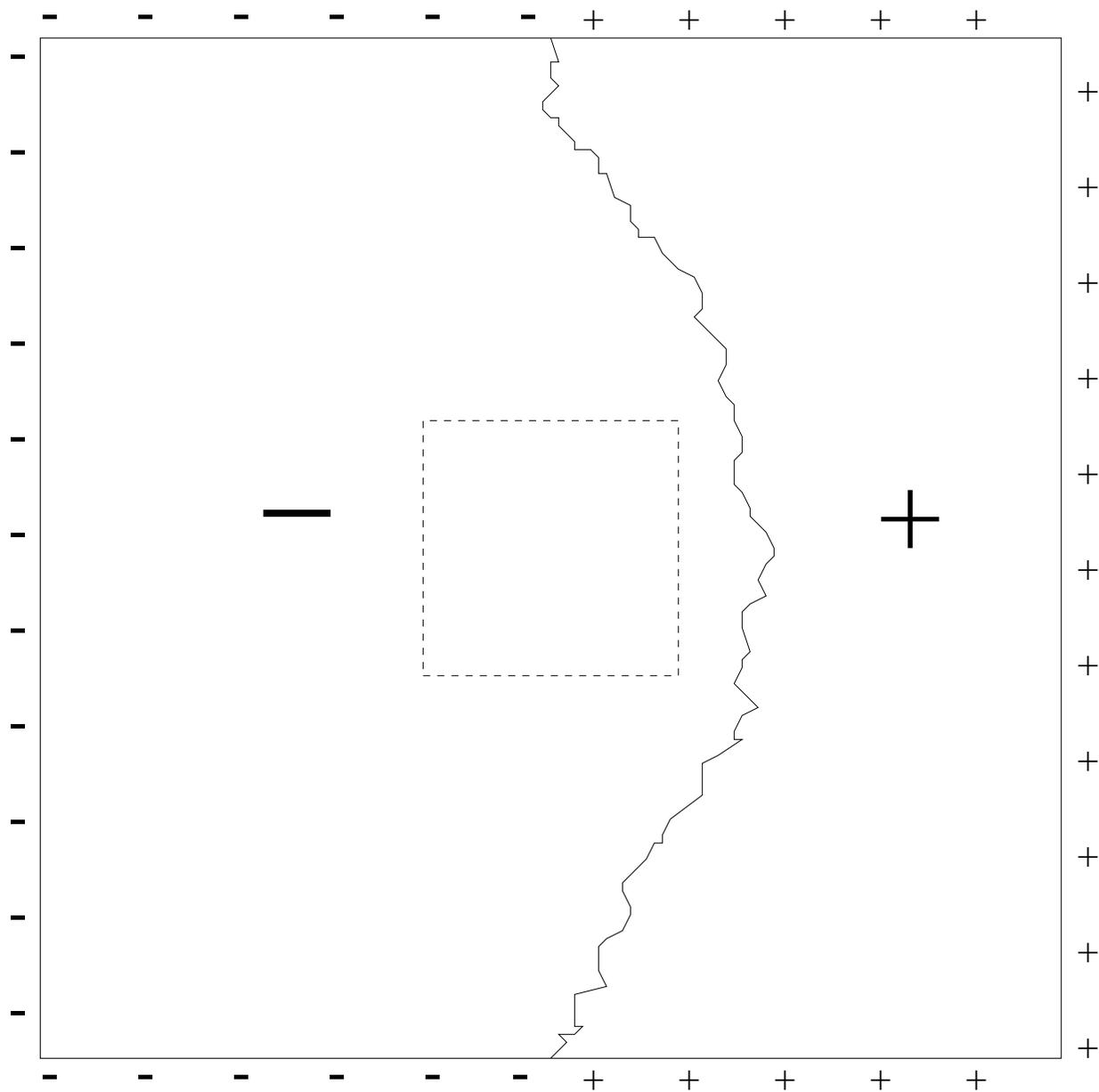
\end{document}